\documentclass[aps,twocolumn,secnumarabic,amsmath,amssymb,nobalancelastpage,nofootinbib,natbib,preprintnumbers]{revtex4-1}
\pdfoutput=1
\usepackage[
      colorlinks=true,
      linkcolor=blue,
      urlcolor=blue,
      filecolor=black,
      citecolor=red,
      pdfstartview=FitV,
      pdftitle={},
        pdfauthor={Paulo Bedaque, Evan Berkowitz, Aleksey Cherman},
        pdfsubject={},
        pdfkeywords={},
        pdfpagemode=None,
        bookmarksopen=true
      ]{hyperref}
      
\usepackage{amsmath, bm, graphicx}
\usepackage{epstopdf}
\usepackage{amscd}
\usepackage{slashed}

\newcommand{\Kplus}{{\ensuremath{K^+}}}

\newcommand{\Kzero}{{\ensuremath{K^0}}}

\newcommand{\fpi}{\ensuremath{f }}
\newcommand{\f}{\fpi}
\newcommand{\pF}{\ensuremath{p_F}}

\newcommand{\M}{\ensuremath{\mathbb{M}}}
\newcommand{\Q}{\ensuremath{\mathbb{Q}}}

\newcommand{\Lg}{\ensuremath{\mathcal{L}}}
\newcommand{\Lge}{\ensuremath{\mathcal{L}_{\text{eff}}}}
\newcommand{\Lgperp}{\ensuremath{\mathcal{L}_\perp}}

\newcommand{\goesto}{\ensuremath{\rightarrow}}
\newcommand{\adjoint}{\ensuremath{{}^\dagger}}

\newcommand{\magnitude}[1]{\ensuremath{\left|#1\right|}}
\newcommand{\del}{\ensuremath{\nabla}}
\newcommand{\grad}{\del}

\newcommand{\Tr}[1]{\ensuremath{\text{Tr}\left( #1 \right)}}
\newcommand{\Det}[1]{\ensuremath{\text{Det}\left(#1 \right)}}
\newcommand{\diag}[1]{\ensuremath{\text{diag}\left(#1 \right)}}
\newcommand{\inverse}{\ensuremath{^{-1}}}
\newcommand{\oneover}[1]{\ensuremath{\frac{1}{#1}}}
\newcommand{\hc}{\text{h.c.}}
\newcommand{\commutator}[2]{\ensuremath{\left[#1,#2\right]}}
\newcommand{\expect}[1]{\ensuremath{\left\langle#1\right\rangle}}

\newcommand{\cross}{\ensuremath{\times}}

\newcommand{\order}[1]{\ensuremath{\mathcal{O}\left(	#1	\right)}}
\newcommand{\omegatilde}{\ensuremath{\tilde{\omega}}}
\newcommand{\ktilde}{\ensuremath{\tilde{k}}}
\newcommand{\nn}{\nonumber}

\begin{document}

\title{Vortons in dense quark matter}
\author{Paulo F. Bedaque$^{1}$}
\email{bedaque@umd.edu}
\author{Evan Berkowitz$^{1}$}
\email{evanb@umd.edu}
\author{Aleksey Cherman$^{1,2}$}
\email{a.cherman@damtp.cam.ac.uk}
\affiliation{$^{1}$ Maryland Center for Fundamental Physics,
Department of Physics, University of Maryland,
College Park, MD 20742-4111 \\
$^{2}$Dept. of Applied Mathematics and Theoretical Physics, University of Cambridge, CB3 0WA, UK}

\preprint{DAMTP-2010-122}
\preprint{UMD-40762-494}

\begin{abstract}
At large baryon number density, it is likely that the ground state of QCD is a color-flavor-locked phase with a \Kzero\ condensate.  The CFL+\Kzero phase is known to support superconducting vortex strings, and it has been previously suggested that it may also support vortons, which are superconducting vortex rings.  We reexamine the question of the stability of vortons, taking into account electromagnetic effects, which make leading-order contributions to vorton dynamics but were not investigated in previous work.  We find that current-carrying and electrically charged vortons can be stabilized either by their angular momentum, by Coulomb repulsion, or by a combination of both effects.  
\end{abstract}

\maketitle

\section{Introduction}
\label{sec:Introduction}

The fate of strongly interacting matter at densities above nuclear matter remains one of the central open problems in the standard model of particle physics. Its qualitative behavior is reasonably well-understood, albeit with important inherent uncertainties
due to the lack of a reliable non-perturbative calculational method in QCD.
 Knowledge accumulated mostly in the last ten years indicates the existence of a color superconducting phase (for  reviews on this vast topic see, for instance \cite{Alford:2001dt,Alford:2007xm,Rajagopal:2000wf}). Color superconductivity arises from the pairing---in the BCS sense---of quarks in a certain combination of spin, flavor and color; different phases arise depending on which quark-quark combinations pair up. At asymptotically high densities the pattern of symmetry breaking due to the quark pairing in three-flavor QCD is known  as weak coupling calculations are reliable \cite{Son:1998uk,Pisarski:1998nh,Schafer:1999jg} in this regime. The symmetry-breaking pairing involves quarks of all colors and flavors and goes by the name of color-flavor-locking (CFL). At lower densities, of the kind possibly found in the interiors of astrophysical compact objects, the problem is more complicated. Many proposals for the phases of QCD at these more phenomenologically-relevant densities have been made in recent years, but a few stand out as appearing more likely based on model calculations.  Among those few phases, the CFL-$\Kzero$ phase\cite{Bedaque:2001p107,Kaplan:2001qk} appears especially likely as its existence is suggested by very model independent arguments  at least in some range of densities\footnote{The main loophole in the arguments is the possibility of an unexpectedly large strength of  $U(1)_{A}$-breaking effects at high density\cite{Schafer:2002ty}.}. An unusual type of solitonic excitation of the CFL-$\Kzero$ phase---the ``vorton''---is the topic of this paper.

The low energy degrees of freedom of the CFL phase are the Goldstone bosons resulting from the symmetry breaking leading to quark pairing. These degrees of freedom have some similarities to the pseudo-scalar nonet found in zero density QCD with one important difference: the ``kaons'' are actually lighter than the ``pions'' \cite{Son:1999cm,Beane:2000ms,Son:2000tu}. The interactions these pseudoscalars exert on one another as well as their interaction with electroweak currents are largely dictated by symmetry considerations.  In fact, the low energy effective theory describing them is very similar to chiral perturbation theory for QCD at zero density\cite{Casalbuoni:1999wu}.  At asymptotically high densities, when the chemical potential $\mu$ is much larger than the quark masses, the quark masses can be neglected. As the density is lowered, the quark masses, in particular the strange quark mass, become more important.  The effect of the strange quark mass can be tracked down to the appearance of a fictitious strangeness chemical potential for the mesons. In particular, the ``$\Kzero$'' and ``$\Kplus$'' mesons become lighter and, at some critical value of the parameter $m_s^2/\mu$, the \Kzero\ meson becomes massless and condenses, leading to the CFL-\Kzero\ phase\cite{Bedaque:2001p107,Kaplan:2001qk}.  Only small isospin breaking effects, like quark masses and electromagnetism, favor the condensation of $\Kzero$ over condensation of $\Kplus$.  In the absence of a $\Kzero$ condensate the $\Kplus$ promptly condenses. Besides the usual vortex solution related to the breaking of the baryon number symmetry $U(1)_{B}$, the CFL-\Kzero\ phase contains another type of vortex related to the breaking of the (approximate) symmetry $U(1)_{s-d}$, the phase of the $\Kzero$ field. At the core of this $\Kzero$ vortex, the $\Kzero$ field vanishes and the $\Kplus$ field may acquire an expectation value. 

The condensation of $\Kplus$ radically changes the properties of the vortex.  Since $\Kplus$ is charged, a nonzero expectation value for $\Kplus$ makes the vortex superconducting. This mechanism is a specific realization of the general mechanism of superconducting strings proposed a long time ago in the setting of relativistic field theory models\cite{Witten:1984eb}. Superconducting strings were extensively studied in  possible realizations in grand-unified theories, since they might be formed in the early universe. (For some reviews of this vast topic, see for instance \cite{vilenkin2000cosmic,Hindmarsh:1994re}). Kaplan and Reddy pointed out that the CFL-$\Kzero$ phase contains all of the ingredients necessary for the formation of superconducting strings\cite{Kaplan:2001hh}. More importantly, there is a natural mechanism for the stabilization of superconducting vortex string \emph{loops}, which are often referred to as ``vortons".  If there is a supercurrent flowing through the loop, there is an associated energy proportional to $\sim R I^2$, where $I$ is the current and $R$ the loop radius. As the radius of the loop changes, the current is not constant; it is actually proportional to $1/R$.  Naively, this is because supercurrents are related to the winding of the phase of $\Kplus$ around the loop, a topological invariant.  As we will see here, the argument becomes more subtle once gauge invariance is taken into account, but the behavior $I\sim R\inverse$ is maintained by the conservation of angular momentum.  
Against the contribution of a supercurrent to the energy of the loop, proportional to $I^2 R \sim 1/R$, one must balance the energy of the vortex due to its tension, proportional to $R$. One then expects there to be some finite value of $R$ that minimizes the energy, and is the stable radius of the vorton. Buckley et al. pointed out that this mechanism did not guarantee the stability of quark matter vortons\cite{Buckley:2002mx}. The reason is that the radius of the vorton should be larger that the thickness of the string. One may think that one can always arrange for the radius to be larger by making a larger current to flow through vorton. However, a large current has the effect of quenching the superconducting condensate and destroying the vorton, a phenomenon well known in the context of cosmic superconducting strings\cite{vilenkin2000cosmic}.  On the other hand, the presence of electric charge in the loop has an anti-quenching effect that also resists the shrinking of $R$ and may stabilize the vorton. Only a more detailed study, including numerical estimates, can tell whether vortons are likely to exist in the CFL-$\Kzero$ phase.  

There are some important differences between vortons in the CFL-$\Kzero$ phase and superconducting cosmic strings besides the trivial difference in energy scales. First, the CFL-$\Kzero$ phase  shields electric fields very weakly, at least at the small temperatures relevant to neutron stars, in contrast to the strong shielding at cosmologically high temperatures usually assumed in the study of cosmic strings. Second, the low energy dynamics of the CFL and CFL-$\Kzero$ phases are described by a {\it non-linear} $\sigma$-model.  As usual, the sigma model is written in terms of a compact variable $\Sigma \sim \exp(i \pi/f_{\pi})$, so that there is maximum value $\sim f_{\pi}$ for the meson fields $\pi$.  Since the meson fields acquire expectation values of roughly the same magnitude as this maximum value, it is unclear whether a linear approximation to the non-linear $\sigma$-model, which was used in previous studies of vortons in the CFL-\Kzero\ phase, will suffice for the stability analysis.   It is the purpose of this paper to present the analysis in the full non-linear sigma model.  Additionally, in previous studies, electromagnetic effects were not fully explored.  Here, we will see that these effects make leading-order contributions to the stability of the vorton.   Of course, this should not be entirely surprising:  the effects of isospin breaking due the quark mass splittings, which are so crucial for the very existence of superconducting strings in the CFL-\Kzero\ phase, are known to be roughly of the same scale as the effects of electromagnetic interactions.


\section{CFL+\Kzero\ phase} 
\label{sec:CFLK0_phase}

When quark masses can be neglected as compared to the chemical potential $\mu$, the high-density ground state of 3-flavor QCD is believed to be the color-flavor locked (CFL) phase.  This phase is characterized\footnote{The CFL condensate shown above is  not gauge invariant.  There are gauge-invariant order parameters characterizing the CFL phase\cite{Rajagopal:2000wf}, but much as in the electroweak sector of the Standard Model, for many purposes it is convenient to discuss the physics using non-gauge-invariant language.} by a condensate of the form\cite{Schafer2000269}
\begin{equation}
	\expect{q^{a}_{L,i}C q^{b}_{L,j}}= - \expect{q^{a}_{R,i}C q^{b}_{R,j}} = \mu^{2}\Delta\	\epsilon^{abZ}\epsilon_{ijZ}
\end{equation}
where $q$ are the quark fields which carry color $a,b=1,2,3$, flavor $i,j=1,2,3$, and helicity $R$ (right-handed), $L$ (left-handed) indices.  The size of the condensate $\Delta$ is referred to as the `gap'.  Such a condensate breaks the symmetry from $SU(3)_{C}\times SU(3)_{L}\times SU(3)_{R}\times U(1)_{B} \times U(1)_{A} \goesto SU(3)_{C+L+R}\times \mathbb{Z}_{2} \times \mathbb{Z}_{2}$.  The first $\mathbb{Z}_{2}$ factor is the remnant of $U(1)_{B}$ that leaves the CFL condensate invariant, while the second $\mathbb{Z}_{2}$ factor is the remnant of the approximate $U(1)_{A}$ symmetry of QCD at high density. There are $18=2*8+2$ broken generators, but $8$ of the would-be Nambu-Goldstone mesons are eaten by the color gauge fields.  As a result, the gluons get masses of order $g_{s}\mu$ ($g_s$ is the strong coupling constant)\cite{Son:1999cm}.   The CFL condensate breaks the vacuum $U(1)_{EM} \subset SU(3)_{L+R}$ gauge symmetry, but a linear combination of $\lambda_{8}$ color generator with the $U(1)_{EM}$ generator annihilates the condensate, generating an unbroken $U(1)_{Q}$ gauge symmetry.   We will henceforth refer to the resulting $U(1)_{Q}$ dynamics as electromagnetism for simplicity\footnote{In fact, at high density, it turns out that the $\lambda_{8}$ part of the generator of it $U(1)_{Q}$ is small when $\alpha_{EM}$ is small\cite{Litim:2001mv}.   Also, the differences between $\alpha_{Q}$ and $\alpha_{EM}$, and the dielectric constant $\epsilon_{\textrm{CFL}}$ in the CFL phase and $\epsilon_{\textrm{vacuum}}$ are both suppressed by the smallness of $\alpha_{EM} \sim 1/137$.}.

The remaining ten broken generators are associated with physical Nambu-Goldstone modes.  Two of these are associated with the superfluidity resulting from $U(1)_{B}$ and $U(1)_{A}$ breaking, and will not be important in what follows.  The other eight form a multiplet transforming under $SU(3)_{C+L+R}$ the same way the zero density pseudoscalar octet transform under flavor $SU(3)$.  At energies small compared to $\Delta$, the low-energy dynamics of the CFL phase is dominated by these physical NG modes.  The dynamics of the NG ``mesons'' that arise in the CFL phase are described by an effective field theory \cite{Son:2000tu,Beane:2000ms,Casalbuoni:1999wu} which bears close resemblance to $SU(3)_{F}$ chiral perturbation theory. 

The effective theory can be written in terms of a chiral field $\Sigma$, which transforms as $\Sigma \rightarrow g_{L}\Sigma g_{R}$ under $SU(3)_{L}\times SU(3)_{R}$.  The eight CFL mesons $\pi^{a}, a=1,\ldots, 8$ appear as fluctuations of the $\Sigma$ field around its vacuum value $\Sigma_{0}$, and are packaged inside $\Sigma$ as $\Sigma=e^{i\pi^{a}\lambda^{a}/\f} \Sigma_{0}$.  Here $\lambda^{a}$ are the standard Gell-Mann matrices with $\Tr{\lambda^{i}\lambda^{j}}=2 \delta^{ij}$, and \f\ is the equivalent of the pion decay constant. 

The lowest-order effective Lagrangian describing the CFL+\Kzero\ phase is~\cite{Son:1999cm,Beane:2000ms,Son:2000tu,Casalbuoni:1999wu,Bedaque:2001p107}
\begin{align}
	\Lge 	&=	\frac{\f^{2}}{4}\Tr{ \grad_{0}\Sigma\adjoint\cdot\grad_{0}{}\Sigma - v^{2}D_{i}\Sigma\adjoint\cdot D_{i}\Sigma}\nonumber\\
			&\phantom{=\ }+2 A \Det{\M}\Tr{\M\inverse \Sigma + \hc} - \oneover{4}F_{\mu \nu}F^{\mu\nu}	\label{eq:lagrangian} ,
\end{align}
where
\begin{align}
	D_{\mu}\Sigma 	&= 	\partial_{\mu}\Sigma - i A_{\mu}\commutator{\Q}{\Sigma},	\nonumber\\
	\grad_{0}\Sigma &=	D_{0}\Sigma + i\commutator{\frac{\M\M\adjoint}{2 \pF}}{\Sigma}, \label{eq:CovariantDerivatives}
\end{align}
\pF\ is the Fermi momentum (which we take to be $\mu$), $A_{\mu}$ is the electromagnetic potential, and $\Q=\frac{e}{3}\diag{2,-1,-1}$ is the quark charge matrix under the $U(1)_{Q}$ symmetry.  $F_{\mu \nu}$ is the electromagnetic field strength tensor, while $\M$ is the mass matrix $\M=\diag{m_{u},m_{d},m_{s}} \approx\diag{2,5,95}$~MeV.   

The $\M^{2}$ dependence of the mass term is different from the mass term encountered in zero-density chiral perturbation theory, which is linear in $\M$.  The reason for this can be traced to the fact that at high densities, the dominant source of chiral symmetry breaking is the CFL condensate rather than the usual chiral condensate.   The CFL condensate preserves left-handed and right-handed quark number modulo two, which forbids mass terms with odd powers of $\M$.  At lower densities, one expects that a $\langle \bar{q}_{R} q_{L}\rangle$ condensate will become non-negligible, leading to the appearance of $\mathcal{O}(\M)$ terms in the effective theory.  Here we assume that the density is large enough that the $\mathcal{O}(\M)$ terms are negligible. The $\mathcal{O}(\M^{2})$ term in eq.~(\ref{eq:lagrangian}) is not the most general  $\mathcal{O}(\M^{2})$ term consistent with the symmetries.  As shown in \cite{Bedaque:2001p107}, however, the other two possible terms are suppressed by a power of $\Delta/\mu$ compared the one in eq.~(\ref{eq:lagrangian}).

In contrast with chiral perturbation theory, the low-energy constants appearing in the CFL effective theory are calculable in the $\mu \goesto \infty$ limit.  At very large $\mu$, one finds~\cite{Son:1999cm}\cite{Son:2000tu}
\begin{equation}
	\f^{2}=\frac{21-8\ln2}{18}\frac{\mu^{2}}{2\pi^{2}},\hspace{.75em}v^{2}=\oneover{3},\hspace{.75em}\text{and}\hspace{.5em}A=\frac{3}{4\pi^{2}}\Delta^{2},
\end{equation}
where $\mu$ is the chemical potential and $\Delta$ is the gap.  

The equation of motion for $\Sigma$ is
\begin{equation}\label{eom-sigma}
	\frac{\f^{2}}{4}\left(	\grad_{0}\grad_{0}\Sigma- v^{2}D_{i}D_{i}\Sigma	\right) = 2A \Det{\M}\ {\M\inverse}\adjoint.
\end{equation}
The equations of motion for $A^{0}$ and $A^{i}$ are
\begin{align}
	\partial_{\mu}F^{\mu0}	&=	-\frac{i f^{2}}{4}\Tr{\grad^{0}\Sigma\adjoint\cdot\commutator{\Q}{\Sigma}-\hc}\nonumber\\
	\partial_{\mu}F^{\mu i}	&=	-\frac{i v^{2} f^{2}}{4}\Tr{D^{i}\Sigma\adjoint\cdot\commutator{\Q}{\Sigma}-\hc}\label{eom-gauge} .
\end{align}
It is straightforward to read off expressions for the electrical charge and current.  
We note that there is a subtlety: a field that has no time dependence still carries charge. In fact, the time-dependence of the fields is not a gauge-invariant concept. Only the combination of time-variation, charge chemical potential and gauge potential is gauge-invariant. 

For values of $\M$  small compared to $\mu$, the vacuum value of $\Sigma$ is simply the unit matrix and the meson mass spectrum can be read off from an expansion around $\Sigma_{0}=1$. This is the CFL phase.  In this phase, the charged pions, charged kaons and neutral kaons have masses~\cite{Bedaque:2001p107}
\begin{align}
	m_{\pi^{\pm}} 	&= 	\sqrt{\frac{4A}{\f^{2}}(m_{d}+m_{u})m_{s}}			,\nonumber\\
	m_{K^{\pm}}		&=	\sqrt{\frac{4A}{\f^{2}}m_{d}(m_{s}+m_{u})}			,\\
	m_{K^{0},\bar{K^{0}}}	&=	\sqrt{\frac{4A}{\f^{2}}m_{u}(m_{s}+m_{d})}	,\nonumber	
\end{align}
which arise from the \order{\M^2} mass term in the effective Lagrangian.  The neutral pion and eta mesons have masses of a similar form, but as they lie on the diagonal when other fields vanish they commute with the $\M\M\adjoint$ matrix in the covariant timelike derivative and thus are never enticed to condense.

As one considers larger values of quark masses, the $[\M\M^\dagger/2p_F,\Sigma]$ terms in the $\grad_{0}$ covariant derivatives become non-negligible, and act as a chemical potential for the mesons, effectively shifting the kaon masses by $\pm m_s^2/2p_F$\footnote{The shift in the pion masses due to the ``strangeness chemical potential'' term is small, of order $\pm (m_d^2-m_u^2)/2p_F$\cite{Bedaque:2001p107}.}.  For  values of $m_s$ larger than a critical value of order $m_s \sim m_{u,d}^{1/3}\Delta^{2/3}$, the neutral kaons mass-squared would be negative. Instead, the neutral kaons condense and the field $\Sigma$ picks an expectation value $\Sigma_{0} \neq 1$.  Since \Kzero\ is a neutral boson, this is a superfluid phase.

The \Kplus\ is also driven to condense by the strangeness chemical potential (assuming the the opposite charges required by charge neutrality are energetically cheap). The question then arises as which field, \Kzero\ or \Kplus, actually condenses. Two factors favor the condensation of \Kzero. First, the small quark mass difference $m_d-m_u$ makes the neutral kaons slightly lighter than the charged ones. Second, a \Kplus\ condensate carries electric charge and, in order to maintain charge neutrality, electrons have to be present. The electrons have an energetic cost due to their mass and kinetic energy up to the Fermi level. For this reason the condensation of \Kzero\ is favored. There is, however, a counteracting effect: fluctuations of the electromagnetic field lower the energy of charged condensates. A full calculation of this effect in the CFL-\Kzero\ phase requires the knowledge of the coefficient of an additional term in the effective Lagrangian for the mesons contributing to the electromagnetic correction to the \Kplus\ mass \cite{Bedaque:2001at}.  For most reasonable values of this coefficient it is the \Kzero\ field that condenses and we will assume in this paper that that is the case. 

With all other fields vanishing, the vacuum expectation value of \Kzero\ condensate is given by~\cite{Bedaque:2001p107}
\begin{equation}
\label{eq:Kzero-soln}
	\cos\left(	\frac{\magnitude{\Kzero}\sqrt{2}}{\f}	\right) = \frac{m_{\Kzero}^{2}}{\mu_{sd}^{2}}
\end{equation}
where $m_{\Kzero}$ is the neutral kaon mass coming from the $\mathcal{O}(\M^{2})$ mass term, while $\mu_{ij} = (m_{i}^{2}-m_{j}^{2})/(2\mu)$ is the effective chemical potential coming from the presence of \M\  matrices in the time-like covariant derivatives.   This expression determines $\magnitude{\Kzero}$ if $\mu_{sd}>m_{\Kzero}$; if  $\mu_{sd} < m_{\Kzero}$, there is no \Kzero\ condensate.


\section{Vortons} 
\label{sec:vortons}
The CFL+\Kzero\ superfluid supports global vortex strings as a consequence of the breaking of $K^{0}$ number.  At the center of such vortices the vacuum expectation value of \Kzero\ vanishes due to the single-valued nature of the field.  As we will see shortly, in the CFL+\Kzero\ phase, the cores of vortices are large enough to be describable within the mesonic EFT. The vanishing of \Kzero\ in the cores of vortex strings provides room for the next lightest boson, the \Kplus\ meson, to condense.  Thus, the cores of the \Kzero\ vortex strings may contain a charged condensate, which is a superconductor.  These vortices are therefore superconducting vortex strings \cite{Witten:1984eb}, and may carry electrical current and charge.  Since the pions are significantly heavier that the kaons we will work in the approximation where the pion fields are turned off. We will keep the $U(1)_{Q}$ gauge fields and the $\Kzero, \Kplus$ fields,  and include all of their non-linear interactions.  Previous studies of vortex strings and vortons in the CFL-\Kzero\ phase did not take into account electromagnetic effects on the dynamics of the strings, and worked only to quartic order in the meson fields for simplicity.   


\begin{figure}[tbp]
	\centering
	\includegraphics[width=\columnwidth]{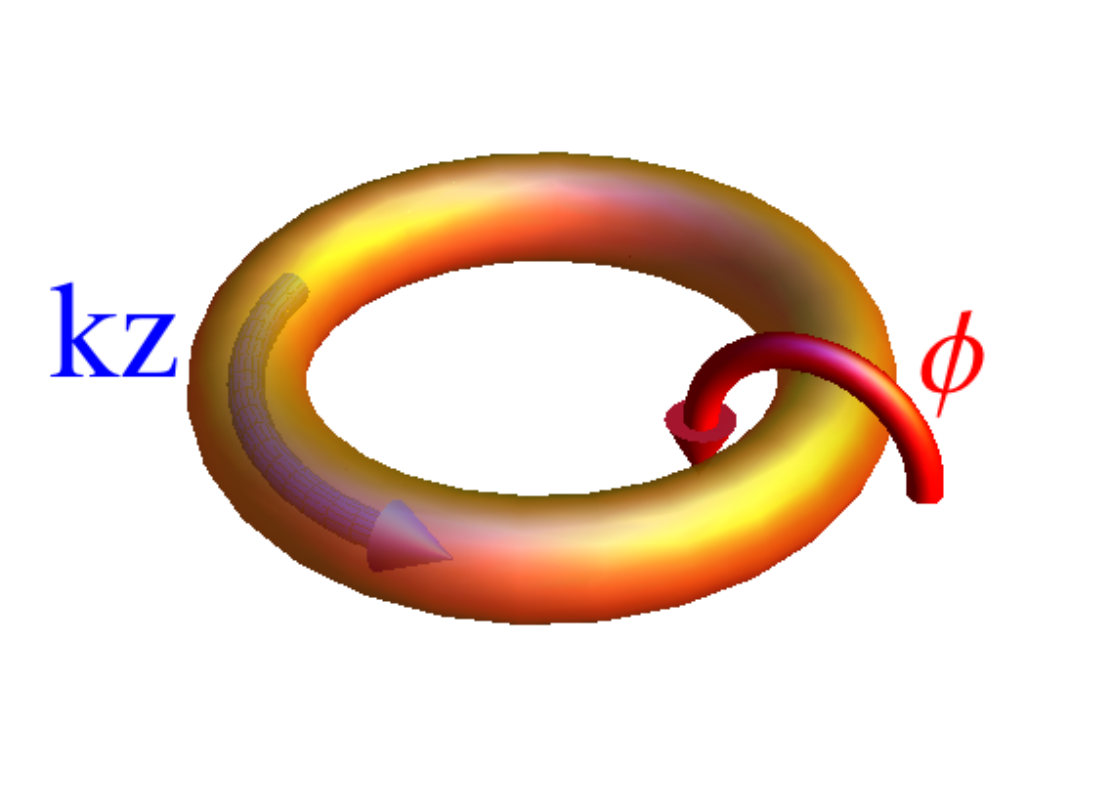}
	\caption{A vorton with arrows indicating the direction of phase change of the \Kzero\ ($\phi$) and \Kplus\ ($k z$)  fields.}
	\label{fig:Rbydelta}
\end{figure}

To describe a single long $\Kzero$ vortex in the $z$ direction the $\Kzero$ field takes the form
\begin{equation} 
\label{eq:KZeroField}
	\Kzero = \Kzero(r) e^{il\phi}.
\end{equation}
where $\phi$ is the axial coordinate, $r$ the axial distance, and $l$ is the vorticity.  For definiteness, we henceforth take $l=1$.  Since the vortex may have an overall electric charge per unit length $\mathcal{Q}_\perp$ and can carry an electrical current $I$, an appropriate general anzatz for the $\Kplus$ field takes the form 
\begin{equation}
\label{eq:KPlusField0}
	\Kplus = \Kplus(r) e^{i(\omega t-k z)}.
\end{equation}
where $k$ is the wavenumber along the $z$ direction, $t$ is time and the frequency $\omega$ contributes to the charge density.  Since a $K^{+}$ field of the form above carries charge and current, it generates an electromagnetic field. The energy of a vortex configuration is higher that the ground state so there is an energetic cost in creating a vortex. The excess of the vortex energy per unit of length is the string tension. The string tension is actually infrared divergent, a well known feature of global vortices. A finite energy is obtained if the vortex exist in a finite container or if the vortex forms a closed loop.  We are interested in the second case, and we will think of the $z$ direction as the direction along the loop.

We now consider vortons, which are closed circular vortex loops.  Generically, the string tension of a vortex loop tends to make such closed strings smaller. On the other hand, if the vorton carries current, the kinetic energy of the \Kplus\ field due to its (covariant) derivative along the z-direction grows as the vorton shrinks. In addition, the presence of charge will help stabilize the vorton due to Coulomb repulsion and, indirectly, by helping stabilize the \Kplus\ condensate supporting the current. There are also magnetic effects to consider as the current on opposite sides of the loop repel each other. Finally, there is a contribution to the energy of a vorton due to weak interactions that, for a very large vorton, can be appreciable. The $U_{ds}(1)$ symmetry related to \Kzero\ number, as well as the  $U_{ds}(1)$ symmetry related to \Kplus number, is explicitly broken by weak interactions. The consequence for the structure of a \Kzero\ vortex is that it is energetically favorable for the \Kzero\ phase  to vary with the angle around the vortex in a inhomogeneous way. In fact, it is favorable for the phase to change little with the angle except in a narrow range of angles where the phase shifts somewhat abruptly by $2\pi$ \cite{Son:2001p108}. The consequence is the formation of a domain wall attached to the vortex carrying some amount of energy. For a closed vortex forming a vorton this wall is a membrane stretching across the vorton (which justifies the name ``drum vortons" sometimes used to describe these objects). The tension of the domain wall is, however, a weak interaction effect and consequently very small. It is only important for very large vortons. As we will see, for the vortons considered in this paper the membrane is a negligible effect.

All the contributions discussed above are important in different regions of  parameter space and should be taken into account. Obviously, finding numerical solutions of the full set of coupled non-linear partial differential equations describing a closed loop is quite challenging.  They are hard to find even in the case of a straight vortex. Fortunately, in order to  get a semi-quantitative understanding of the physics, one does not have to solve the full 3D problem.  

Before tackling the problem of determining the meson and electromagnetic field solutions,  let us discuss the conserved quantities constraining the time evolution of vortons. The electric charge of a vorton can only change through  the absorption of an electron and the conversion of a \Kplus\ into a \Kzero. The rate of this weak process is suppressed by the fact that, as we will find out, typical vortons occupy a volume much smaller than the size of the electron clouds around them. There will be a time scale, however, where this process will be equilibrated. For times larger than that scale the electric charge of the vorton will be driven to zero, the \Kplus\ condensate will also vanish and the vorton will collapse. Our current study is valid for shorter time scales. We leave as a future project to estimate the time scale for charge equilibration. A similar statement can be made about the angular momentum of the vorton. There are processes (quasi-particle and/or photon emission) that can change the angular momentum of the vorton. We will consider a time scale shorter than that, where the angular momentum of the vorton can be considered to be constant.  There is one more fixed quantity: the winding number of the $\Kplus$ field along the vorton.

If  the \Kplus\ field is to be periodic in $z$ while remaining single-valued, $k$ in eq.~(\ref{eq:KPlusField0}) must be quantized so that
\begin{equation}
	2\pi N = \int_{0}^{2\pi R}dz\, k	\label{topology}
\end{equation}
where $N$ is an integer and $R$ is the vorton radius, so that $k= N/R$.  The winding number $N$ cannot change smoothly, since the topology forces it to be an integer;  it can only change by tunneling.   One expects the tunneling rate to be very small, suppressed by the exponential of the energy it costs to set the \Kplus\ condensate to zero everywhere along the vorton, which is what is necessary to change $N$.  Therefore, one expects $N$ to be a conserved quantity.  This intuition is correct.  However, as we now discuss, specifying $N$ does not fix the properties of the vorton.

The quantization of $k$ was already discussed in Ref.~\cite{Buckley:2002mx}, which did not treat electromagnetic effects.  These effects are somewhat subtle.  The time-dependent phase of the $\Kplus$ field in eq.~\eqref{eq:KPlusField0} can be changed by $U(1)_{Q}$ gauge transformations, and is thus is not an observable.  However $N = k R$ remains a physical observable, since it cannot be changed by periodic gauge transformations.  In fact, given a loop of superconducting wire, the quantization of the fluxoid connects $N$ to the magnetic flux $\Phi_{B}$ that threads the superconducting loop and the current density $j^{z}$ at the loop's center by \cite{tinkham2004introduction}
\begin{equation}\label{eq:fluxoid}
	2\pi N = e \Phi_{B} - \oint dz \frac{j^{z}(r=0)}{\frac{f^{2}v^{2}}{2e}\Tr{|\commutator{\Q}{\Sigma}|^{2}}} .
\end{equation}
In most familiar condensed matter systems, the superconducting sample is large compared to the London penetration length $1/m_{\gamma}$ so the current that flows through the center of the loop is exponentially small.  Then the integral term in eq.~(\ref{eq:fluxoid}) is negligible and it then appears as though flux itself is quantized.  While qualitatively the flux and fluxoid are different entities, quantitatively they are very similar in this `thick-wire' case.  In our case, we will find that the penetration length is comparable to the thickness of the wire; the vortex is in neither the ``thick" or ``thin" wire regime.\footnote{Cosmic strings can typically be analyzed in the much simpler ``thin" wire approximation \cite{Witten:1984eb}}  Thus, for us the distinction between the fluxoid and the flux is quite significant.

Next, even though in principle $A_{z}$ and $N/R$ (or more properly $\Phi_{B}$ and $N$) are independent physical observables, they do not enter calculations separately. Instead they appear in just such a combination so that they always give the current density $j^{z}$;  this can be traced to the fact that the contributions of $A_{z}$ and $N/R$ both originate in the $U(1)_{Q}$ covariant derivative $D_{z} = \partial_{z}+A_{z}$.  This poses a puzzle, because although $N$ is conserved, it does not determine the current, and we cannot deduce how the current should change as the radius of the vorton changes.  If we increase $N$ by 1 and the magnetic flux by $2\pi/e$, the resulting vorton will have the same current, and so the Lagrangian will not differ for these two different vortons.

In the ungauged case the dependence of the current density $j$ on the vorton radius $R$ is quite straightforward: $j\sim k\sim N/R$. This scaling was, in fact, used to argue for the stability of the ungauged vorton.  Once we take the gauge fields into account, it is not immediately obvious that this stabilizing effect will persist. Conceivably, the gauge field might adjust in such a way as to cancel the $R\inverse$ behavior.  As will be made explicit in section~\ref{sec:stability}, a stabilizing effect of this sort survives, and shows up through the requirement of angular momentum conservation.  

We have argued that in order to describe a vorton, we should specify {\it both} $N$ and the magnetic flux --- these quantities together give the current.  If we specify both the current and the charge, we can compute the angular momentum.  Instead, we will just specify the angular momentum and the charge.  Thus, when we specify the angular momentum, we are actually discussing a whole class of vortons whose magnetic flux and winding number conspire to give the specified value of the angular momentum.  While vortons supporting different amounts of magnetic flux and winding number are in principle physically distinct objects,  the stability and equilibrium properties of vortons related by having the same angular momentum must be the same.

Thus, we are interested in minimizing the action at a fixed charge and angular momentum.  One way to accomplish this is to use Lagrange multipliers $\nu$ and $\Omega$ to define
\begin{equation}
	F= \nu Q + \Omega J - L_{eff}.
\end{equation}
Extremizing $F$ is equivalent to solving the equations of motion arising from the Lagrangian $L$ while guaranteeing that the field configuration supports the correct amount of charge and angular momentum.  We will need to eliminate these Lagrange multipliers appropriately in favor of given values of the charge $Q$ and angular momentum $J$. 

\begin{figure}[tbp]
	\centering
	\includegraphics[width=\columnwidth]{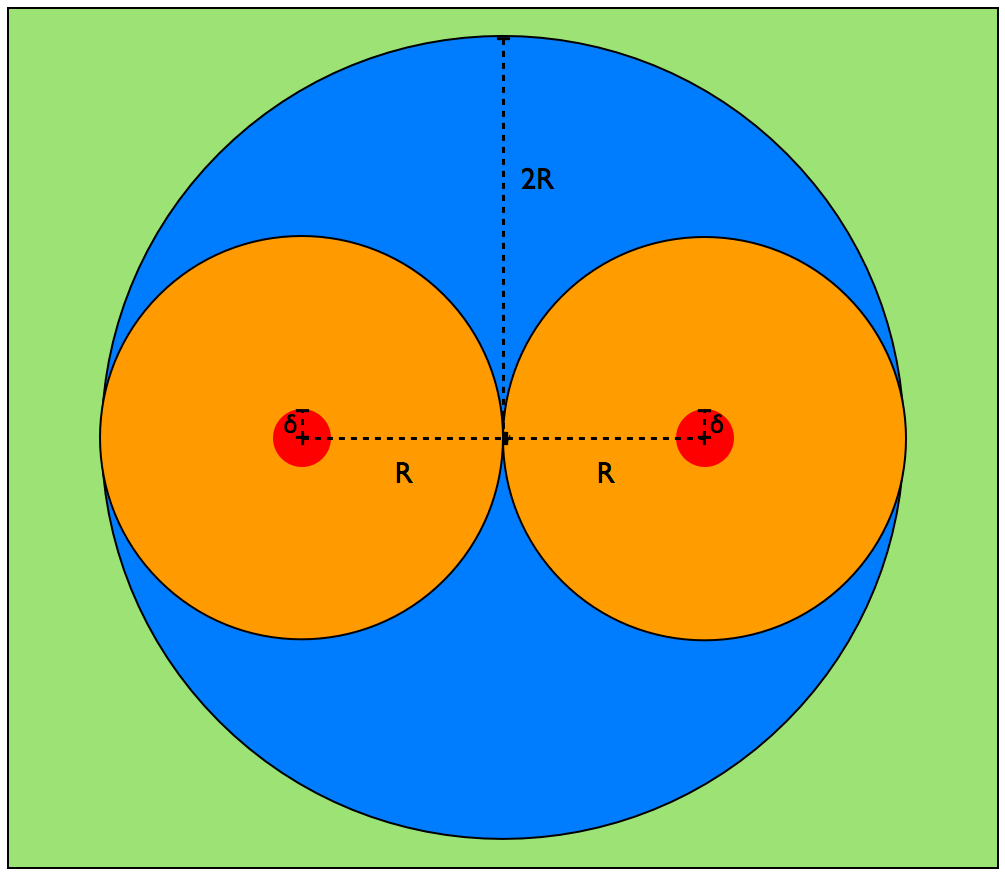}
	\caption{A cross section of the three regions of integration described in the text. The inner torus (orange online) is the region where the energy is approximated by $2\pi R$ times the tension. Outside the sphere (green online) the fields are approximated by the multipole expansion.  The apple-core shape (blue online) is neglected.}
	\label{fig:geometry}
\end{figure}
  
  The calculation of $F$ by integrating its density over space is very complicated due to the geometry of the vorton. We will obtain an approximate solution to this problem by dividing the space into three regions. One is given by a torus of radius $R$ around the vorton (see Figure~\ref{fig:geometry}). The second is the region outside a sphere of radius $2R$ centered on the vorton center. The third region is the small ``apple-core" shaped region between the torus and the sphere. By picking the sphere's radius to be $2R$, we ensure that it just touches the torus but that they do not overlap.
The integral over the torus will be approximated by $2\pi R$ times the quantity
\begin{align}
\mathcal{F}_\perp &= \int_0^R d^2r_\perp (\nu \mathcal{Q}+\Omega \mathcal{J}-\Lg_{eff}),	\nonumber\\
				&=	\nu \mathcal{Q}_{\perp} + \Omega \mathcal{J}_{\perp} - \int_{0}^{R} d^2r_\perp\ \Lg_{eff}
\end{align}  where $ \mathcal{Q}$ and $  \mathcal{J}$  are the  electric charge and  angular momentum {\it densities}, and $\mathcal{Q}_{\perp}$ and $\mathcal{J}_{\perp}$ are the corresponding linear densities.  The quantity $\mathcal{F}_\perp $ plays the role of the string tension when charge and angular momentum are fixed. We will compute $\mathcal{F}_\perp $ from the definition above by taking the fields to be the one in a {\it straight} vortex. This approximation becomes exact as the vorton loop radius $R$ is much larger than the vortex thickness $\delta$. 
The integration over the region outside the sphere is simple. The charge density will vanish in the ansatz for the kaon fields we will adopt (and it is very small in the exact solution). The only contributions to the angular momentum will come from long-distance electric and magnetic fields, treated in the multipole expansion.  The ``apple-core"-shaped region between the torus and the sphere will be neglected. In order to proceed we will now consider the calculation of $\mathcal{F}_\perp$ for a straight vortex.

\subsection{Straight superconducting vortices}
\label{sec:straight_vortices_in_cfl_kzero}

In the presence of a single, infinitely long straight $\Kzero$ vortex in the $z$ direction, the $\Kzero$ field takes the form
\begin{equation} 
\label{eq:KZeroField}
	\Kzero = \Kzero(r) e^{il\phi}.
\end{equation}
where $\phi$ is the axial coordinate, and $l$ is the vorticity.  For definiteness, we henceforth take $l=1$.  Since the vortex may have an overall electric charge per unit length $\mathcal{Q}_\perp$ and can carry an electrical current $I$, an appropriate general anzatz for the $\Kplus$ field takes the form 
\begin{equation}
\label{eq:KPlusField}
	\Kplus = \Kplus(r) e^{i(\omega t-k z)}.
\end{equation}
where $k$ is the wavenumber along the $z$ direction, $t$ is time and the frequency $\omega$ contributes to the charge density.  Since a $K^{+}$ field of the form above carries charge and current, it generates an electromagnetic field.  The equations of motions following from eq.~(\ref{eq:lagrangian})  then give four coupled non-linear ODEs that determine $\Kzero(r), \Kplus(r), A_{t}(r),A_{z}(r)$.

Unfortunately, these equations cannot be solved analytically.   Nevertheless, we can get some insight into the behavior of the solutions by posing a simple form for the radial profiles of the kaon fields  and calculating the electromagnetic response to these background profiles.  On general grounds, one expects that the magnitude of the \Kzero\ condensate in a single-vortex solution will interpolate between the limiting values $\Kzero(0)=0$ and $\Kzero(r\rightarrow \infty) = \magnitude{\Kzero}$ which takes the value given by eq.~(\ref{eq:Kzero-soln}).  At the same time, the \Kplus\ condensate profile will go from some fixed value $\Kplus = \magnitude{\Kplus}$ at $r=0$ to zero at large $r$.  The simplest profiles for the kaon fields that incorporate these general features is
\begin{align}
	\Kzero(r)&=\magnitude{\Kzero}\theta(r- \delta),	&	\Kplus(r) &= \magnitude{\Kplus}\theta(\delta-r) ,\label{eq:steps}
\end{align}
where we will refer to $\delta$ as the string thickness and treat it as a variational parameter alongside the parameter \magnitude{\Kplus}.
 Obviously, these simple step-function ans\"atze will not solve the kaon equations of motion, but we expect them to capture the qualitative features of the full solutions.  While somewhat simplistic, these square profiles provide the ability to solve the electromagnetic equations of motion and understand their back reaction on the kaon fields analytically.
 
Given the above ans\"atze we can solve the electromagnetic equations of motion in closed form.  In keeping with our anzatz, we do not allow the electromagnetic fields to backreact on the {\it shapes }of the kaon fields, but we do allow the electromagnetic effects to change the {\it size} of the charged condensate at the center of the vortex.  In solving for the profiles of the gauge fields, it is useful to note that $k, \omega$ and $A_{t}, A_{z}$ and $\mu_{su}$ are not separately gauge-invariant.  They appear in the equations of motion through the gauge-invariant combinations
\begin{align}
	\omegatilde(r) &= \omega-e A_{0}(r) -\mu_{su}	&	\ktilde(r) &= k + e A_{z}(r).
\end{align}
Alone, the time dependence $\omega$, the size of the gauge field $A_{0}$, and the quantity $\mu_{su}$ are meaningless: explicit time dependence and the chemical potential may be absorbed into an overall shift in $A_{0}$ and only grouped into $\omegatilde$ do they take on a gauge-invariant meaning.  An analogous statement may be made about $k$ and $A_{z}$. In the static solution we are looking for we may assume that the gauge fields are time independent.  Then, in Coulomb gauge and with the ans\"atze in eqs. \eqref{eq:KZeroField}, \eqref{eq:KPlusField} and \eqref{eq:steps}, the gauge field equations of motion \eqref{eom-gauge} reduce to
\begin{align}
	\nabla^{2}\omegatilde	&=	\omegatilde m_{\gamma}^{2}\; \theta(\delta-r)\nonumber\\
	\nabla^{2}\ktilde	&=	\ktilde v^{2}m_{\gamma}^{2}\; \theta(\delta-r),	\label{eom-gauge-reduced}
\end{align}
where $m_{\gamma}^{2}= \f^{2} e^{2} \sin^{2}(\magnitude{\Kplus}\sqrt{2}/\f)$ is the mass of the photon inside the vortex core (outside of the vortex core the photon is massless).  We also see that the photon mass is bounded by $m_{\gamma}^{2}=e^{2} \f^{2} $, which is an effect that would be missed if one expanded $\Sigma$ in the kaon fields.  The charge and current densities may be read off directly from the right side of these equations. Notice that the charge and current depend on the value of the electromagnetic fields themselves.

The electromagnetic field equations of motion, by gauge invariance, are homogeneous equations of \omegatilde\ and \ktilde\ respectively.  In contrast, when electromagnetism is sourced by spin-half fermions, the equations of motion are not homogeneous---the electric charge is proportional to the fermion number alone.  Instead, we find that a nonzero \Kplus\ field only carries charge if the gauge field is nonzero.  This implies that if the charge density vanishes at the wire's core there cannot be electric charge anywhere.  This is apparent in the explicit solutions to eq. \eqref{eom-gauge-reduced}.  An analogous statement holds for current density.

The solutions to the equations of motion are
\begin{align}
	\omegatilde(r)	&= 	\omegatilde_{0}\ I_{0}( m_{\gamma} r)	&	 r<\delta\nonumber\\
					&=	\omegatilde_{0}\ \left(	I_{0}(m_{\gamma}\delta) + m_{\gamma}\delta I_{1}(m_{\gamma}\delta)\log\left(	\frac{r}{\delta}	\right)	\right)	&	r>\delta\label{omega-soln}\\
	\ktilde(r)		&=  \ktilde_{0}\ I_{0}(v m_{\gamma}  r)		&	r<\delta\nonumber\\
					&=  \ktilde_{0}\ \left(	I_{0}(vm_{\gamma}\delta) + vm_{\gamma}\delta I_{1}(vm_{\gamma}\delta)\log\left(	\frac{r}{\delta}	\right)	\right)	&	r>\delta\label{k-soln}
\end{align}
where $I_{0}$ and $I_{1}$ are the 0$^{\text{th}}$ and 1$^{\text{st}}$ modified Bessel function and $\omegatilde_{0} \equiv \omegatilde(r=0)$ and $\ktilde_{0} \equiv \ktilde(r=0)$. 
The values of $\omegatilde_{0}$ and $\ktilde_{0}$ are fixed by the boundary conditions. We can relate them to the charge and current flowing through the string. In fact,
using the known large-$r$ behavior of the electromagnetic fields due to an infinitely-long charged current-carrying wire, we find that the charge per unit length $\mathcal{Q}_\perp$ and the total current $I$ carried by the vortex are given by
\begin{align}
	\mathcal{Q}_\perp &= +\frac{2\pi}{e} \omegatilde_{0} m_{\gamma}\delta I_{1}(m_{\gamma}\delta)
	= \omegatilde_{0} m_{\gamma}\delta f_1(m_\gamma\delta),	\label{lambda-eqn}\\
	I	&= -\frac{2\pi}{e} \ktilde_{0} v m_{\gamma} \delta I_{1}(vm_{\gamma}\delta)
	= -\ktilde_{0} v m_{\gamma}\delta f_1(v m_\gamma\delta)
\end{align} We emphasize that, due to gauge invariance, the solutions only depend on $\omega$, $\mu_{su}$ , $k$ , $A_0$ and $A^z$ through the combinations $\tilde\omega$ and $\tilde k$. A change in, say, $\omega$ is absorbed into a change in $A_0$ in order to keep $\tilde\omega$ the same.  The apparent difference in sign between the charge and current is due to the difference in sign of the gauge fields when defining $\omegatilde$ and $\ktilde$.

We now calculate the Lagrangian per unit of vortex length   \Lgperp\  for the gauge fields in eqs.~(\ref{omega-soln},\ref{k-soln}).  The integral in the radial direction is infrared divergent due to the logarithmic behavior of the gauge field at large distances. 
As discussed above, we will only need to integrate the Lagrangian up to a radial distance $R$.
The Lagrangian per unit length
\Lgperp\ is then a function of  $\{\magnitude{\Kplus}, \omegatilde_{0}, \ktilde_{0}, \delta, R\}$ defined by 

\begin{equation}
 	\Lgperp =  \int_{0}^{R} d^{2}r_{\perp}\ \Lge.
\end{equation} A calculation with the ansatz in eq.~(\ref{eq:steps}) and the solutions in eqs.~(\ref{omega-soln}, \ref{k-soln}) gives a somewhat unwieldy but analytic expression
\begin{align}\label{eq:Lperp_ansatz}
\Lgperp &=
\omegatilde_0^2 m_\gamma \delta f_3( m_\gamma\delta)
-\ktilde_0^2 v m_\gamma \delta f_3(v m_\gamma\delta)
+
A f_5,
\end{align} where
\begin{align}
f_3(x) &=      \frac{\pi}{e^2}
\left(
I_0(x)I_1(x) + x I_1^2(x) \log(R/\delta)
\right),
\end{align}
and
\begin{align}
f_5 &= 4\pi \delta^{2} \left(m_{s}m_{u}+m_{d}(m_{s}+m_{u})\sqrt{1-\left(	\frac{m_{\gamma}}{f e}	\right)^{2}}	\right)	\nonumber\\
	&+4\pi (R^{2}-\delta^{2})\left(	m_{d}m_{s}+m_{u}(m_{d}+m_{s})\left(	\frac{m_{\Kzero}}{\mu_{sd}}	\right) 	\right)	\nonumber\\
	&+\frac{\pi f^{2}}{2A}\left(	R^{2}-\delta^{2}	\right)\mu_{sd}^{2}	\left(1-\left(	\frac{m_{\Kzero}}{\mu_{sd}}	\right)^{2}\right)\nonumber\\
	&-\frac{\pi f^{2}}{A}\left(l^{2}v^{2}\log\left(	R/\delta	\right)\right)\left(1-\left(	\frac{m_{\Kzero}}{\mu_{sd}}	\right)^{2}\right).
\end{align} The dependence on \Kplus\ is hidden inside the photon mass $m_\gamma = e f \sin(\sqrt{2} K^+/f)$.  The first two contributions to $f_{5}$ come from the integral of the mass term in the Lagrangian.  The second two come from the kinetic \Kzero\ terms. Incidentally, we mention that there are electromagnetic one-loop radiative corrections contributing to the energy split between a \Kzero\ and a \Kplus\ condensate\cite{Bedaque:2001at}. The magnitude of this effect is comparable to the one given by the difference in quark masses included above. Unfortunately, the calculation of its precise value  depends on the value of a counterterm in $\mathcal{L}_{eff}$ not yet computed even in QCD perturbation theory. For reasonable values of this counterterm this contribution is not sufficient to revert the roles played by \Kzero\ and \Kplus. In what follows, we will disregard this contribution.

Extremizing \Lgperp\ with $\omegatilde_{0}, \ktilde_{0}$ fixed leads to an equation of motion for $\magnitude{\Kplus}$.  Depending on the values of $\omegatilde_{0}, \ktilde_{0}$ two kinds of solutions are possible.  One class of solutions has a nonzero vacuum expectation value \Kplus\  and is associated with the existence of  a charged condensate at the core of the vortex which breaks the $U(1)_{Q}$ symmetry.  The other class of solutions has a quenched condensate, satisfying  $\sin(\sqrt{2}\magnitude{\Kplus}/\f)=0$, and does not break $U(1)_{Q}$.  These other insulating solutions correspond to $\Sigma=\diag{-1,1,-1}$, which is diagonal and so commutes with \Q.  This implies that the photon mass vanishes, as can be seen from eq.~\eqref{eom-gauge}.   The existence of this peculiar quenched solution is a direct result of the kaon magnitude being a compact variable, and is not seen upon linearization.  Despite this odd property of the full non-linear theory, it remains the case that a large wavenumber generically causes quenching. This is because one can show that if the preferred value of $\sqrt{2}\magnitude{\Kplus}/\f$ is near 0, increasing $\ktilde_{0}$ tends to push the condensate smaller; if the preferred value is near $\pi$, increasing $\ktilde_{0}$ tends to push $\sqrt{2}\magnitude{\Kplus}/\f$ toward $\pi$.

Part of the dependence of the solutions on $I$ and $\mathcal{Q}_\perp$ is easy to understand and was already anticipated in previous work which used a linearized version of eq.~(\ref{eq:lagrangian})~\cite{Buckley:2002ur,Buckley:2002mx}.  A large $|\ktilde_{0}|$, which enables a large current, simultaneously decreases the magnitude of the charged condensate and eventually leads to quenching, while a large \magnitude{\omegatilde_0}, which helps increase $\mathcal{Q}_\perp$, helps stabilizes the condensate.

If our main interest were the straight vortex we could now minimize \Lgperp\ in relation to \magnitude{\Kplus} and $\delta$ and trade the dependence on $\omegatilde_0$ and $\ktilde_{0}$ by $\mathcal{Q}_\perp$ and $I$. That would tell us whether or not the condensate is quenched inside the vortex and how thick the vortex is in a situation where the charge density and current are held fixed.  This is, however, a very different problem from the one we aim to solve. In a vorton with changing radius it is the total charge and angular momentum that is held fixed, not the charge density and current. For this reason, we will proceed to calculate the remaining terms contributing to $\mathcal{F}_\perp$.

The charge per unit length is given by the expression in eq.~(\ref{lambda-eqn}). The angular momentum per unit of length requires a little more work.
The total momentum density for the field configuration is
\begin{equation}
	T^{0i}=\frac{\f^{2}}{4}\Tr{vD^{i}\Sigma\adjoint\cdot\grad^{0}\Sigma + \hc} + \left(\vec{E}\cross\vec{B}	\right)^{i}.
\end{equation}
We are interested in the momentum that flows along the vortex.  With our ans\"atze, we find
\begin{align}
	T^{0z}	&=	\omegatilde_{0} \ktilde_{0} v \f^{2} \sin^{2}\left(	\frac{\magnitude{\Kplus}\sqrt{2}}{\f}	\right)\times\\
			&	\left(I_{0}(m_{\gamma}r)I_{0}(m_{\gamma}v r)+I_{1}(m_{\gamma}r)I_{1}(m_{\gamma}v r)\right)	&r<\delta\nonumber\\
			&	\frac{\delta^{2}}{r^{2}}I_{1}(m_{\gamma}\delta)I_{1}(m_{\gamma}v\delta)	& r>\delta\nonumber,
\end{align}
The contribution from outside of the vortex is provided entirely by the electromagnetic Poynting vector.  The total angular momentum per unit length is given by 
\begin{align}\label{angmom-eqn}
	\mathcal{J}_\perp &= \int_{0}^{R} d^{2}r_\perp\ \sqrt{R^{2}+r_{\perp}^{2}+2 R r_{\perp}\cos\phi}\  T^{0z} \\
	                                  &\approx\int_{0}^{R} d^{2}r_\perp\ R\  T^{0z} \nn\\
	                                  &= R m_\gamma\delta  \ktilde_0 \omegatilde_0 f_2,
	                                  \end{align}
	   with
	                                
    \begin{align}
	                               f_2 =&   -\frac{2\pi v }{e^2 (1+v)}
	                                  (
	                                 I_0(v m_\gamma\delta )  I_1(m_\gamma\delta )
	                                 +
	                                   I_0( m_\gamma\delta)  I_1(v m_\gamma \delta)\nn\\
	                                 &+
	                                   (1+v)m_{\gamma}\delta  I_1( m_\gamma\delta )  I_1(v m_\gamma\delta ) \log(R/\delta)
	                                  ),
	\end{align}
where we have inserted the lever-arm from the vorton's center to the vortex's center, $R$, to approximate the angular momentum per unit length.  Corrections due to different portions of the vortex being different distances from the vorton's center are negligible in the limit $R\gg \delta$ since most of the angular momentum is carried inside of the vorton ($r<\delta$) and not by the electromagnetic field outside of the vorton as we numerically verified.

\subsection{Far fields}
Outside a sphere of radius $2R$ the main contribution to $L$ and $J$ come from the electric and magnetic fields. We will approximate the fields in this region by their multipole expansion: the electric field by the monopole contribution and the magnetic field by its dipole form.  The fields are
\begin{align}
	\vec{E}_{out} &= 	\frac{Q}{4\pi r^{2}}\hat{r}	\\
	\vec{B}_{out} &=	\oneover{4\pi r^{3}}\left(	3 (\vec{m}\cdot\hat{r})\hat{r}-\vec{m}	\right)
\end{align}
$Q=2\pi R \mathcal{Q}_{\perp}$ is the total vorton charge, and $\vec{m} = \pi R^{2}\ I\ \hat{z}$ is the magnetic moment of the vorton, assuming it lies in the $xy-$plane.

Their contribution to the total Lagrangian outside of the sphere is
\begin{align}\label{eq:far_energy}
L_{out}	&=	\int_{outside} d^3r \mathcal{L}_{eff} = \int d^3r \frac{\vec{E}_{out}^2-\vec{B}_{out}^2}{2} \nn\\
		&= 	\frac{1}{2}\int_{2R}^\infty dr r^2 \ d\Omega\	\vec{E}_{out}^2-\vec{B}_{out}^2	\nonumber\\
		&=	\frac{Q^{2}}{16\pi R} - \frac{m^{2}}{96 \pi R^{3}}\nonumber\\	
  		&= 	\left(2\pi R\ \delta m_{\gamma}	\right)\left(	\omegatilde_0^2\ f_{7} - v\ktilde_{0}^{2}\ f_{8}	\right)
\end{align} 
where we have grouped various factors to appear similar to \eqref{eq:Lperp_ansatz} when we evaluate $2\pi R \Lgperp$, so that
\begin{align}
f_7 &=  \frac{\pi^{2}\ m_{\gamma}\delta}{2\ e^{2}}I_{1}(m_{\gamma}\delta)^{2},\nn\\
f_8 &=  \frac{\pi^{2}\ v m_{\gamma}\delta}{48\ e^{2}}I_{1}(v m_{\gamma}\delta)^{2}.
\end{align}
The difference between $f_7$ and $f_{8}$, aside from the factors of $v$ that are associated with $\ktilde_{0}$, is the dimensionless factor that arises from the fact that the magnetic field is a dipole and not a monopole like the electric field.

Further contributions from multipole terms will be down by at least factors of $4$, which correspond to the factor of 2 in front of the cutoff $r_{min}=2R$ as well as geometrical factors.  The higher electric and magnetic multipole moment contributions will have the effect of changing the coefficients of electric monopole and magnetic dipole terms at \order{1}, but will not change the qualitative dependence on $R,Q$, and $J$ shown above. It may seem strange that the contributions of the electric and magnetic energy have opposite signs as both should be repulsive and help vorton stabilization, and thus presumably should have the same sign when they enter $F$. This impression is an artifact of the way we set up our calculation up to now. After the Lagrange multipliers $\nu$ and $\Omega$ are eliminated in favor of the total charge both electric and magnetic energy will have the same repulsive effect.

The angular momentum also has a contribution coming from the far fields as the Poynting vector $\vec{S}=\vec{E}_{out}\cross \vec{B}_{out}$ field lines forms closed circles around the vorton. This contribution can be calculated as
\begin{align}
J_{out} 	&= \int_{outside}d^{3}r\ \vec{r}\cross\vec{S}	\nonumber\\
					&= \int_{2R}^{\infty} r^{3}dr\ d\Omega\ \vec{r}\cross\left(	-\frac{Q m \sin\theta}{(4\pi)^{2}r^{5}}	\hat{\phi}\right)	\nonumber\\
					&= \frac{Q \vec{m}}{12\pi R}	\nonumber\\
					&= (2\pi R) R m_\gamma \delta \tilde k_0 \tilde\omega_0\ f_6,\label{eq:jout}
\end{align} with
\begin{equation}
f_6 =  - \frac{\pi^{2}v m_{\gamma}\delta}{3\ e^{2}} I_{1}(m_{\gamma}\delta)I_{1}(v m_{\gamma}\delta),
\end{equation}
arranged to be easily added to the contribution to the angular momentum from $2\pi R \mathcal{J}_{\perp}$.

\section{Vorton stability}\label{sec:stability}

We can now collect the results in eqs.~(\ref{lambda-eqn},\ref{eq:Lperp_ansatz},\ref{angmom-eqn},\ref{eq:far_energy}) and \eqref{eq:jout} and obtain the function to be minimized $F$ by adding the contributions coming from the three regions of space (see Figure~{\ref{fig:geometry}}).
$F$ is a function of $\{\magnitude{\Kplus}, \omegatilde_{0}, \ktilde_{0}, \delta, R,\nu, \Omega\}$ but not $Q$ or $J$---these quantities are expressed wholly terms of those variables as well. 
We find
\begin{align}
	F =&\ \nu\left(2\pi R \omegatilde_{0} m_{\gamma}\delta f_{1}(m_{\gamma\delta})\right) + \Omega( 2\pi R^{2}m_{\gamma}\delta \ktilde_{0}\omegatilde_{0} )(f_{2}+f_{6})	\nonumber\\
	&- 2\pi R \delta m_{\gamma}\omegatilde_{0}^{2}\left(	f_{3}(m_{\gamma}\delta) + f_{7}	\right)	\nonumber\\
	&+ 2\pi R \delta m_{\gamma} v \ktilde_{0}^{2}\left(	f_{3}(v m_{\gamma}\delta)+f_{8}	\right) \nonumber\\
	&- 2\pi R A f_{5}
\end{align}
We now minimize $F$ in relation to  $\omegatilde_{0}$ and $\ktilde_{0}$. This gives us two relations that we can use to eliminate $\nu$ and $\Omega$:

\begin{align}
\Omega &= -\frac{2 \ktilde_{0} v}{\omegatilde_{0} R }\frac{f_{3}(v m_{\gamma}\delta)+ f_{8}}{f_{2}+f_{6}},\nn\\
\nu &= 2 \omegatilde_{0}\left(	\frac{f_{3}(m_{\gamma}\delta)+f_{7}}{f_{1}}+\frac{\ktilde_{0}^{2} v}{\omegatilde_{0}^{2}} \frac{f_{3}(v m_{\gamma}\delta)+f_{8}}{f_{1}}	\right)
\end{align}

Plugging these values of $\nu$ and $\Omega$ back into $F$ we find
\begin{align}
F =&\ 2\pi R m_\gamma \delta
\left[
\omegatilde^2 \left(	f_{3}(m_{\gamma}\delta)+f_{7}	\right) + \ktilde^2 \left(	f_{3}(v m_{\gamma }\delta) + f_{8})	\right)\right] \nonumber\\
 	&- 2\pi R\ A f_5
\end{align}
 Having eliminated the Lagrange multipliers from $F$, we can trade $\omegatilde_{0}$ and $\ktilde_{0}$ for the conserved quantities.  The quantities we are given by
\begin{align}
	Q &= 2\pi R\ \mathcal{Q}_{\perp} &&= 4\pi^{2} R \omegatilde_{0} m_{\gamma}\delta I_{1}(m_{\gamma}\delta),	\\
	J &= 2\pi R\ \mathcal{J}_{\perp} + J_{out}	&&= 2\pi R^{2}m_{\gamma}\delta \omegatilde_{0}\ktilde_{0}(f_{2}+f_{6})
\end{align} 
and so we can eliminate $\ktilde_{0}$ in favor of $J/\omegatilde_{0}$ and $\omegatilde_{0}$ in favor of $Q$.  Thus, the conserved quantities that enter into $F$ always appear in the combinations  $Q$ and $J/RQ$.  
 
At this point we have $F$ as a function of the total charge $Q=eZ$, the angular momentum $J$, the vorton radius $R$, the string thickness $\delta$ and the charged condensate at the center of the vortex \magnitude{\Kplus}. We still need to minimize $F$ in relation to $R$, $\delta$ and \magnitude{\Kplus} while keeping $Q$ and $J$ fixed.
 Unfortunately, the expressions cannot be minimized analytically.  We attack the question of the stabilization of the vorton numerically.  We fix $Z$ and $J$ at some given values, and then numerically minimize $E$ with respect to $\magnitude{\Kplus}$, $\delta$, and $R$ simultaneously.

\begin{figure}[htbp]
	\centering
	\includegraphics[width=\columnwidth]{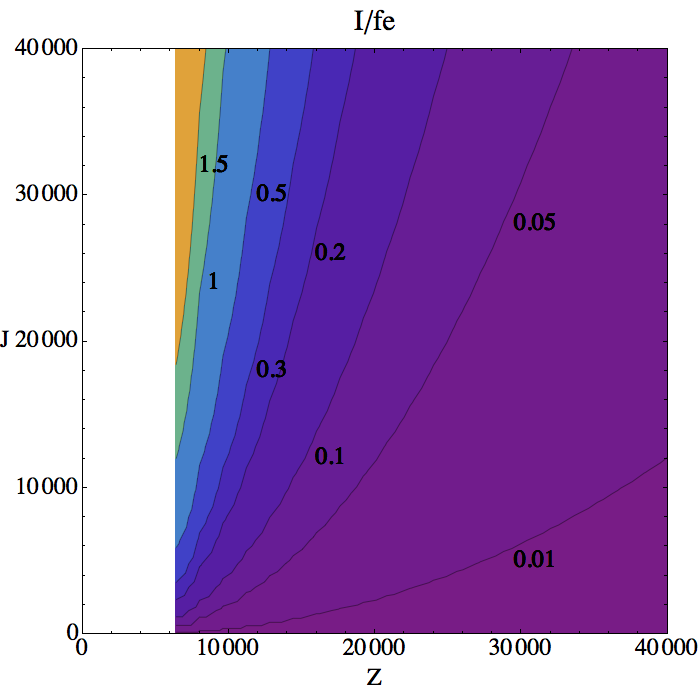}
	\caption{The current $I$ flowing in a vorton of charge $Ze$ and angular momentum $J$ at its preferred radius.}
	\label{fig:totalcurrent}
\end{figure}

\begin{figure}[htbp]
	\centering
	\includegraphics[width=\columnwidth]{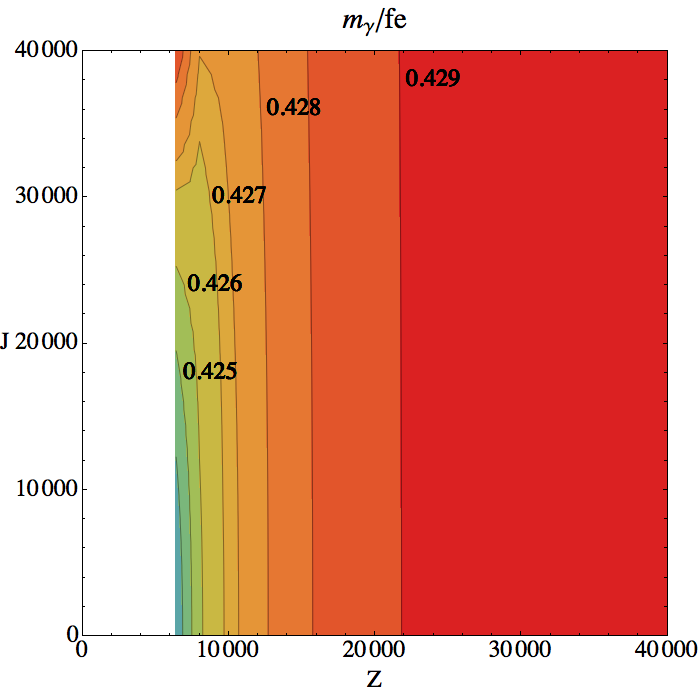}
	\caption{The photon mass for a vorton at its preferred radius with given charge $Ze$ and angular momentum $J$.}
	\label{fig:photonmass}
\end{figure}

For definiteness, we will show results for a gap $\Delta=66$ MeV, a chemical potential $\mu=450$ MeV.  We plot the resulting equilibrium radii in units of the string thickness, $R_{0}/\delta$ in Figure~\ref{fig:Rbydelta} as a function of $Z$ and $J$.  The photon mass normalized by the maximum possible photon mass $m_{\gamma}/\f e$ is shown in Figure~\ref{fig:photonmass}, while and the total electrical current $I/\f e$ flowing in the vorton is shown Figure~\ref{fig:totalcurrent}, all as functions of $Z$ and $J$.

\begin{figure}[htbp]
	\centering
	\includegraphics[width=\columnwidth]{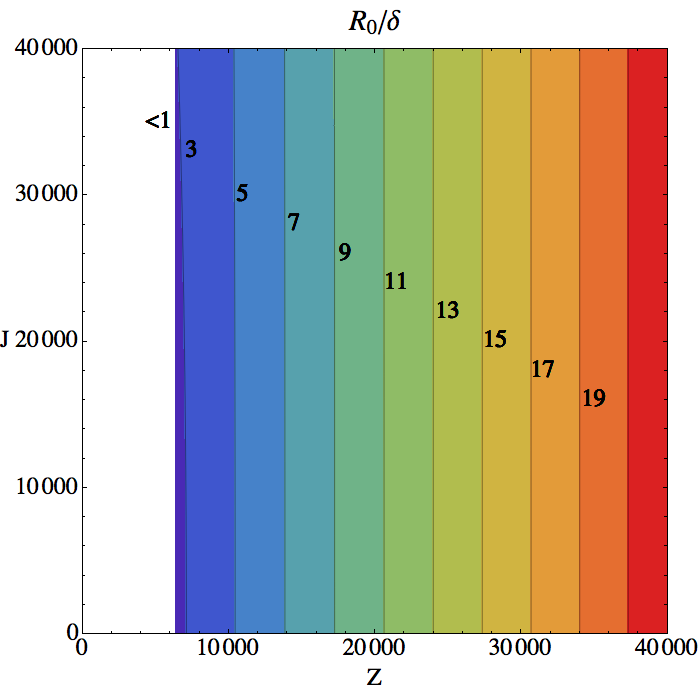}
	\caption{The equilibrium radius $R_{0}$ compared to the string thickness $\delta$ at equilibrium as a function of charge $Ze$ and angular momentum $J$. In all plots the region with $R_0<\delta$ are shown in white.}
	\label{fig:Rbydelta}
\end{figure}

While the string thickness is a variational parameter and is varied for each ($Z,J$) pair, its change is very slight---for this choice of $\Delta$ and $\mu$ and quark masses, the value of $\delta$ is varies from 25 to 29 fm.  This small change is  unimportant for understanding the presented figures---one may think of $\delta$ as a fixed parameter.  It is important to recognize, however, that  $\delta$ sets the scale for distance over which the fields vary appreciably. The numerical value we found is larger than the scale $1/\Delta\approx 4$ fm at which the effective theory in eq.~(\ref{eq:lagrangian}) would break down. This justifies {\it a posteriori} the use of the effective theory in analyzing the physics of vortons.

In Figure~\ref{fig:Rbydelta} we see that the equilibrium radius of a vorton increases with $Z$ but is relatively independent of $J$.  This is an artifact of choosing $J$ as the vertical axis: the angular momentum always enters expressions divided by $Z$ as noted above.  However, these axes are useful for seeing the other expected behavior.  For example, in Figure~\ref{fig:totalcurrent} we see the current that flows through equilibrium-sized vortons.  Since at this scale, the radius is relatively constant with changing $J$, a higher angular momentum must directly imply a higher current.  Such behavior is evident in Figure~\ref{fig:totalcurrent}: for a fixed $Z$ the current rises as $J$ increases.

From Figure~\ref{fig:totalcurrent} we see that larger $Z$ lead to smaller currents. On the other hand, larger $Z$ correspond to larger $R_0$ on account of Coulomb repulsion between opposite sides of the vorton (and also seen on  Figure~\ref{fig:Rbydelta}). Thus there is an inverse relation between current and radii: $I \sim 1/R$. This is the scaling one would naively expect if we assumed the current to be proportional to $N/R$ as in the ungauged case. Qualitatively, this repulsive effect is still there in the more rigorous analysis presented here.

\begin{figure}[htbp]
	\centering
		\includegraphics[width=\columnwidth]{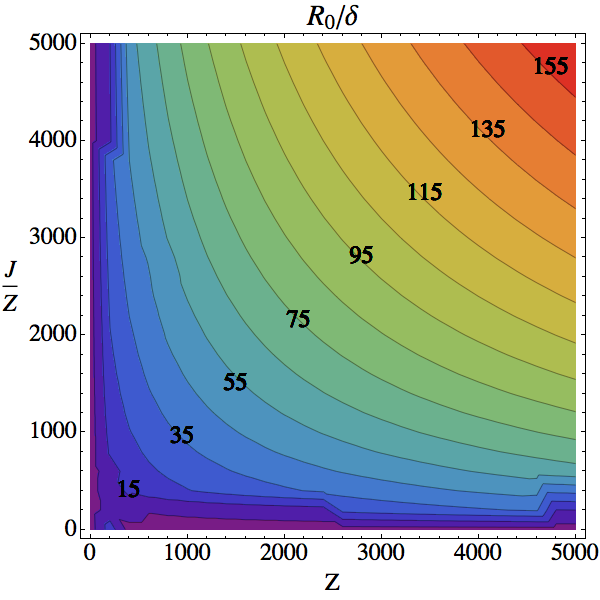}
	\caption{The equilibrium radius $R_{0}$ compared to the string thickness $\delta$ replotted with the vertical axis rescaled by $Z$.}
	\label{fig:Rbydelta-rescaled}
\end{figure}

To see the angular momentum barrier at work, we need to zoom out so that the suppression of $J$ by $Z$ is negligible.  To this effect, in Figure~\ref{fig:Rbydelta-rescaled} we plot $R_{0}/\delta$ while fixing the variables $Z$ and $J/Z$ for the same choice of $\Delta$ and $\mu$.  One clearly sees the tendency for vortons to grow as $J/Z$ is increased.  

From this numerical example, typical among other reasonable values of the parameters, we learn some important lessons about vorton stability. The first and most important one is that in order to have a vorton that really looks like a torus with $R\gg \delta$ charges of the order of $Z$~ many thousands are required. Second, even for these large values of $Z$ the equilibrium radii are small enough that the energy from the membrane stretched across the vorton contributes negligibly to the energy budget. As mentioned before $\delta$ seems large enough that the effective mesonic theory is believable. A more delicate issue is the charge neutrality of the medium as a whole.

Given the requirement of overall charge neutrality for a physical system like a neutron star core, one expects there to be light charged leptons (either electrons or positrons) present in the bulk to cancel the electric charge of the vortons.  Thus, electrons will orbit positively charged vortons. If the orbit size is much larger than the vorton size $R$, their orbits will have a typical size of the order of the Bohr radius  $a_0$ divided by $Z$: $R_e\approx a_0/Z$. For $Z\approx 10000$, the orbit size is $\approx 5$ fm, smaller even than the thickness $\delta$ of the vorton. This means that electrons will likely shield appreciably the electric force helping to keep the vorton stable, a contribution not included in our calculation. On the other hand, only a fraction of the electron will be close enough to the vortons to shield the electric force. Just like in a regular atom, the exclusion principle pushes some of the electrons to much bigger orbits and they are very much less attracted to the vortons as the other electrons shield them from the effect of the vorton charge. 
One simple way to include some of the Pauli principle effects is to use the Thomas-Fermi approximation that, at least for a point nucleus, gives a typical electron orbit of the order of $R_e\approx a_0/Z^{1/3} \approx 2300$ fm (for $Z=10000$), which is much larger than the vorton radius $R\approx 120 $ fm. This suggests that only a small fraction of the electron orbiting a vorton will effectively shield the repulsion between different parts of the vorton.
The proximity of the scales discussed above, however, should serve as caution against hastily dropping the electron shielding effect. A better study of this problem would requires us to understand the ``atomic physics" of the vortons, where the toroidal shape of the ``nucleus" and the electromagnetic fields sourced by the vorton have a significant effect on the electron orbitals. We will postpone this analysis to a future paper. If it turns out that the electron orbits are indeed much larger than the vortons, then ``vortonic chemistry" will include all the molecules we would find on Earth if elements with $Z\approx 10000$ were stable.

There is another class of potentially stable vortons which is not affected by the electron shielding. For small $Z$ but very large angular momentum, say  $J \approx 10^5$, vortons can be stabilized at radii $R\gg \delta$. Since for smaller charges even the tightest  electron orbits will be larger than the vorton and, in any case, electrons are less effective at shielding magnetic forces, the electron cloud cannot destabilize the vortons.

\section{Discussion} 
\label{sec:comments}
We have shown that the CFL+\Kzero\ phase supports stable charged current-carrying vortons, supported either by conservation of angular momentum, Coulomb repulsion, or a combination of both effects, and have estimated their size as a function of the relevant parameters as charge and angular momentum. Besides a more accurate calculation of the vorton structure, a number of improvements and future directions suggest themselves at this point.

However, one crucial step is missing for connecting all of these speculations with phenomenological consequences for neutron stars.   Unfortunately, despite some effort, we were unable to understand and quantify the possible mechanisms of vorton formation during the early stages of neutron star evolution. For this reason we can not, at the moment, estimate their density.  But all possible phenomenological consequences of the existence of vortons on the physics of neutron stars depend crucially on the vorton density. In reference \cite{Kaplan:2001hh} a possible mechanism for the generation of vortons was suggested, including the generation of a supercurrent. Still, no numerical estimate of the vorton density was made.  Also, the crucial role of the charge in stabilizing the vorton was not understood at the time  \cite{Buckley:2002mx} and no mechanism for depositing charge along the vorton string is presently known. Still, if vortons are to exist in neutron star cores, it seems likely that they would have to be produced in non-equilibrium processes soon after the formation of the neutron star.  If current and charge carrying vortons are indeed formed, the evidence we presented above strongly suggests they would be stable against strong and electromagnetic processes.   Estimating the number of vortons produced at the birth of the neutron star seems to us to be the most important problem to be addressed before further progress can be made in the topic. Another essential element missing in our analysis is an estimate of the vorton lifetime against weak processes, in particular the electron capture of inner electrons in a vortonic atom that can lower the charge and eventually destabilize the vorton itself.

 One may also imagine that a neutron star cored filled with vortons and electrons might possess crystalline or metallic structure due to the interactions mediated by the electron cloud or through magnetic dipole interactions. At the core of these vortonic atoms there is also a ``nuclear structure" to be understood. We have analyzed vortons in isolation and not considered either their interaction or their stability against fission into two vortons.
Once a bulk with vortons is made, they should be locked in, supported by their conserved quantities and prevented from disappearing, except by weak effects and the possibility of fusing or fissioning.  In the long term, then, one might expect all vortons to either decay or combine toward making the vortonic equivalent of $^{56}$Fe. Most likely, the ``chemical equilibrium" will never be reached, just like the chemical equilibrium among ordinary nuclei is never reached in the early universe.

\acknowledgments 

We thank S. Reddy, D. Kaplan and T. Cohen for discussions and N. Yamamoto for pointing out some misstatements in previous versions of the manuscript.  This research was supported by the U.S. Department of Energy under grant \#DE-FG02-93-ER40762.

\bibliography{vortons}

\end{document}